\documentclass[letter,twocolumn]{jpsj3}
\usepackage{txfonts}

\usepackage{siunitx}
\usepackage{textcomp}
\usepackage{algpseudocode}
\usepackage{algorithm}
\usepackage{here}

\usepackage{tabularx}

\usepackage{graphicx}
\usepackage{float}
\usepackage{textcomp}
\usepackage{amsmath}
\usepackage{listings}
\usepackage{url}

\usepackage{physics}

\newcommand{\f}{f^{}}
\newcommand{\fd}{f^{\dagger}}

\newcommand{\Gu}{\Gamma_{3u}}
\newcommand{\Gv}{\Gamma_{3v}}
\newcommand{\Gs}{\Gamma_{1}}

\newcommand{\FRAC}[2]{{\textstyle \frac{#1}{#2}}}

\newcommand{\EQ}{\! = \!}

\newcommand{\msm}[1]{\mspace{-#1mu}}
\newcommand{\tot}{\mathrm{tot}}

\newcommand{\TK}{T_{\mspace{-3mu} K}}

\title{Numerical Renormalization Group Study of Quadrupole Kondo Effect with the Crystal-field Excited State}

\author{Yuki Kaneko\thanks{kaneko-yuki314@issp.u-tokyo.ac.jp} and Hirokazu Tsunetsugu\thanks{tsune@issp.u-tokyo.ac.jp}}
\inst{The Institute for Solid State Physics, The University of Tokyo, Kashiwa, Chiba 277-8581, Japan
} 

\abst{We have studied the quadrupolar Kondo effect for an impurity 
in cubic environment
with taking account of a singlet excited state $\Gamma_1$, 
which models a $\mathrm{Pr}^{3+}$ ion with a non-Kramers double 
ground state $\Gamma_3$.  
We have used the numerical renormalization group approach 
and determined the phase diagram 
with varying quadrupole Kondo coupling $J$ and 
$\Gamma_1$ excitation energy $\Delta$.  
Two phases are found and identified 
as local Fermi liquid and non-Fermi liquid. 
This non-Fermi liquid phase is characteristic to the two-channel 
Kondo impurity, and a similar phase diagram has been 
also found in other extended quadrupole models.  
We have analyzed in detail the $J$-dependence of 
the Kondo temperature $T_K$
near the phase boundary 
$J_c (\Delta) $  
and found a nice scaling behavior 
with an stretched exponential form 
$T_K \sim \delta J^{\alpha} \exp 
(-\mathrm{const.}/\sqrt{\delta J})$
where $\delta J \equiv J - J_c (\Delta)$.
This differs from the standard scaling form 
and indicates that one needs to consider 
renormalization of multiple coupling constants.
}

\begin{document}
\maketitle

The multi-channel Kondo effect was first discussed by Nozi\`eres and Blandin \cite{Nozieres1980} 
to take account of orbital degrees of freedom into the Kondo effect.
The effect of multiple channels may lead to the non-Fermi liquid (NFL) ground state 
in contrast to the local Fermi liquid (LFL) state \cite{nozieres1974} in ordinary Kondo systems. 
Since then, 
many studies have been performed on the multi-channel Kondo effect, 
and its singular behavior has been theoretically elucidated 
using various approaches such as numerical renormalization group (NRG) \cite{wilson1975,Bulla2008,Pang1991}, 
conformal field theory \cite{Affleck1990,Affleck1991a,Affleck1991b,Affleck1991c}, 
and Bethe ansatz \cite{Andrei1984,Tsvelick1985,Schlottmann1995}. 
However, it is not easy to find model materials modeling a two-channel Kondo effect, 
hindered by the instability of the NFL state against channel asymmetries \cite{Cragg1980,Pang1991,Affleck1992}.
Thus, the design of physical systems realizing such states is an important challenge in strongly correlated systems; 
as well as experimental verification of NFL state.

For realizing the two-channel Kondo effect,
D.~L.~Cox proposed the use of a quadrupole moment 
of an ion with $f^2$-electron configuration such as 
$\mathrm{U}^{4+}$ or $\mathrm{Pr}^{3+}$, 
and this is called the quadrupolar Kondo effect. 
The Kondo coupling works in the orbital degrees 
of freedom, while conduction electron spins 
$\sigma \in \{ \uparrow , \downarrow \}$ corresponds 
to the channel degrees of freedom.
The double-fold degeneracy of their ionic ground state 
is a nonmagnetic one of non-Kramers type 
and guaranteed when their crystal field has a high symmetry 
\cite{Cox1987,Cox1988,Cox1988a}. 
Recent experiments reported a few behaviors suggesting 
the quadrupolar Kondo effect in the 1-2-20 compounds 
PrV$_2$Al$_{20}$ \cite{Sakai2011} and 
PrIr$_2$Zn$_{20}$ \cite{Onimaru2011,Onimaru2016}. 
However, large intersite interactions modify 
their physical properties from the behaviors of the 2-channel impurity model. 
They are candidate materials realizing the quadrupolar Kondo lattice (QKL), 
and several of their properties are consistent 
with the theoretical predictions for the QKL's \cite{Tsuruta2015}.
Furthermore, a recent experiment reported 
the impurity quadrupole Kondo effect in the diluted Pr system 
Y$_{1-x}$Pr$_{x}$Ir$_2$Zn$_{20}$ \cite{Yamane2018}.

One should note one important point about the orbital symmetry 
of an impurity quadrupole in a lattice.  
We consider here the case of cubic lattice.  
Despite a discrete symmetry of the crystal field, 
the quadruple has the continuous O(2) orbital symmetry,  
if the $\Gamma_3$ doublet alone is considered. 
Taking into consideration the singlet $\Gs$ excited state, 
this elevated symmetry is reduced to Z$_3$, which is equivalent 
to the lattice symmetry.  
As for quadrupole lattices, it was found that this 
Z$_3$ orbital anisotropy plays an important role 
of stabilizing various types 
of quadrupole orders \cite{Hattori2014}.  
However, this effect has been neglected 
in most of the theoretical studies on the quadrupole Kondo effect, 
except for a few studies which have investigated their effects on the NFL state 
for a quadrupole model in either tetragonal or hexagonal crystal field.  
They used the NRG method \cite{Koga1995,Koga1996,Koga1997} or 
a variational approach \cite{Kusunose1998} 
and found a non-vanishing part of the parameter space 
where the NFL state is stable. 
This contrasts to the instability of the NFL fixed point 
against an infinitesimal channel anisotropy.  

In this paper, we study the quadrupole Kondo effect with a special 
attention on the effect of orbital Z$_3$ anisotropy.  
To this end, we consider an impurity 
with the excited $\Gamma_1$ singlet and ground-state $\Gamma_3$ doublet 
$\{ \ket{\Gamma_{3u}},\ket{\Gamma_{3v}}, \ket{\Gamma_1} \}$.  
Our model differs from those used in the previous studies.
Koga and Shiba used a realistic model considering 
the total angular momentum $j_z = 5/2$ of conduction electrons \cite{Koga1995}.
Their model is quite complicated and also has a lower symmetry
since they considered hexagonal or tetoragonal crystal field. 

We will study an extended quadrupole Kondo model (EQKM) 
which includes the singlet excited state $\Gamma_1$.
Its NRG Hamiltonian reads 
\begin{subequations} 
\begin{align}
H_N &= \Lambda^{1/2} H_{N-1} + 
\sum_{\alpha \sigma} 
\bigl( \fd_{N-1,\alpha \sigma } \f_{N \alpha \sigma } 
+ \mathrm{H.c.} \bigr), 
\label{eq:eqkmnrg}
\\[-6pt]
\Lambda^{1/2}    H_0 &= 
J \sum_{\sigma} 
\Bigl[ 
\bigl( \fd_{0 u \sigma} \f_{0 v \sigma} + 
       \fd_{0 v \alpha} \f_{0 u \sigma} \bigr) \, Q_x 
\nonumber\\[-10pt]
&\hspace{1cm} 
+ \bigl( n_{0 u \sigma} - n_{0 v \sigma} \bigr) \, Q_z 
\Bigr] + \Delta \ket{\Gamma_1}\bra{\Gamma_1}  , 
\label{eq:eqkmh0}%
\end{align}
\label{eq:HamNRG}%
\end{subequations}
where 
$\Delta$ is the excitation energy of the $\Gamma_1$ state. 
The chemical potential is set to 0 corresponding to the half-filled 
conduction band, and $H_N$ has the partile-hole symmetry.
$\fd_{n \alpha \sigma }$ is the creation operator of 
the electron in an orbital 
$\alpha \msm3 \in \msm3 \{u,v \}$ with spin 
$\sigma \msm3 \in \msm3 \{ \uparrow , \downarrow\}$ 
at the $n$-th site in Wilson chain \cite{wilson1975} 
and 
$n_{0 \alpha \sigma}^{} \msm2 = \msm2 
\fd _{0 \alpha \sigma} \f _{0 \alpha \sigma} $.  
$\Lambda = 3$ is the scaling factor at each RG step.  
The impurity quadrupole operators are 
defined as 
$Q_z = 
 \ket{\Gu}\bra{\Gu}
-\ket{\Gv}\bra{\Gv} 
+ \bigl( a \ket{\Gs}\bra{\Gu} + \mathrm{H.c.} \bigr)$ 
and 
$Q_x = 
\bigl( a \ket{\Gs} -\ket{\Gu} \bigr) \bra{\Gv}
+ \mathrm{H.c.}$ 
with $a=\sqrt{35}/2$, and 
they constitute a basis set of 
the $\Gamma_3$-irreducible representation (irrep)
of the cubic point group.   

We now discuss the symmetry of $H_N$.  
At each NRG step, 
we decompose the Hilbert space into subspaces 
using conserved quantities 
and diagonalize $H_N$ in each subspace. 
Once $H_0$ includes 
the $Q_{z,x}$'s matrix elements involving $\Gamma_1$,  
$H_N$ loses the O(2) symmetry in the $Q_x$-$Q_z$ space, 
and thus its generator,  
$Q_y^{\mathrm{tot}} = 
i (\ket{\Gu}\bra{\Gv} + \sum_{n \sigma} \fd _{n u \sigma}\f_{n v \sigma}) 
+ \mathrm{H.c.}$,  
becomes non-conserved, 
while the SU(2) spin rotation symmetry is unaffected.  
Thus, for subspace decomposition, 
we use 
the pair of quantum numbers 
$\{ C \} = (C_{\uparrow}, C_{\downarrow}) $,  
where 
$ C_\sigma \msm2 = \msm2 
\sum_{0 \le n \le N} \sum_{\alpha=u,v} 
( \fd_{n \alpha \sigma} \f_{n \alpha \sigma} \msm2 - \msm2 1/2)$ 
is the electron number for the spin direction $\sigma$ 
minus $(N \msm2 + \msm2 1)$.  
In our NRG calculations, 
we retained about 1620 total
states at each iteration $N$.  
We have also used a special trick to reduce numerical errors. 
Due to SU(2) spin symmetry, 
the states with $S^{\text{tot}} > 0$ are necessarily degenerate 
among the subspaces $-S^{\tot} \le S_z^{\tot} \EQ 
\FRAC12 (C_{\uparrow} \msm2 - \msm2 C_{\downarrow}) \le S^{\tot}$.
Thus, after diagonalization at each iteration, 
we reset the corresponding eigenenergies by their average value.  
This drastically improved the stability of NRG procedure.  
Without this trick, the system sometimes flows towards 
a false fixed point.

\begin{figure}[tb]
  \begin{center}
      \includegraphics[scale=0.36]{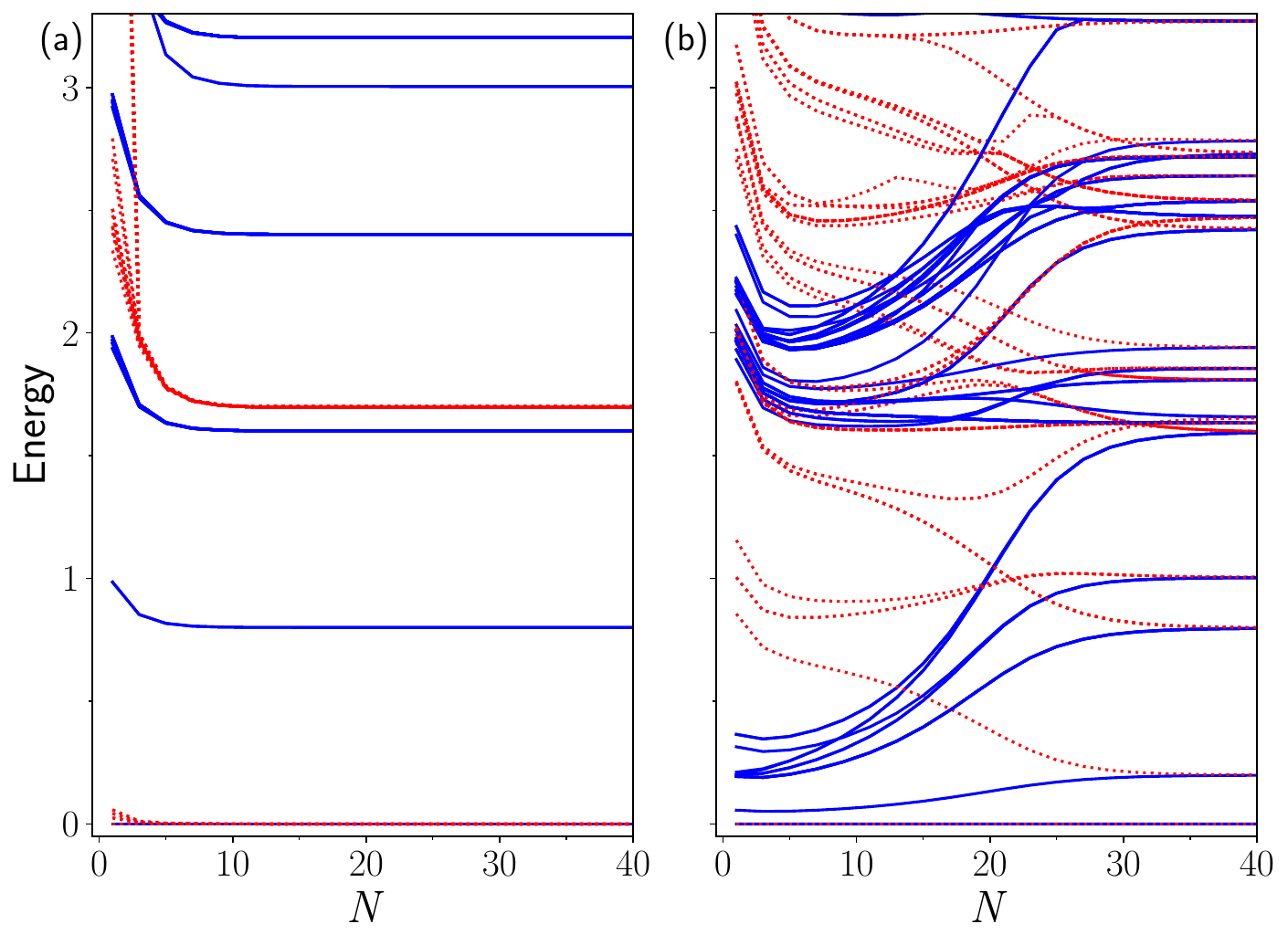}
      \caption{(Color online) 
      NRG energy flow of low-lying levels for  $(J,\Delta)$= (a) (1.0, 1.0) and (b) (0.1, 10). 
      Data are shown by solid blue lines for even $N$ and 
      dotted red lines for odd $N$. 
      $\Lambda \EQ 3$.}
    \label{fig:flow}
  \end{center}
  \end{figure}

Identification of the phase requires the analysis of the RG fixed point.   
For each $(J,\Delta)$-point, we start from $H_0 $ and repeat 
NRG iterations until the low-energy spectrum converges. 
Figure \ref{fig:flow}(a) plots the RG energy flows 
for $(J,\Delta) \msm2 = \msm2 (1.0,\, 1.0)$ 
and this is the behavior of the LFL fixed point.   
The ground state is singlet with $\{ C \} \EQ (0, 0)$. 
The spectrum can be described by quasi-particle excitations, 
and thus the levels are equally separated by 
the single-particle energy $\eta_1 \EQ 0.80041$.
For even $N$, the first excited multiplet consists of the states with 
$\{ C \} \EQ (\pm 1, \, 0)$ or $(0, \, \pm 1)$, and 
each $\{ C \}$-state is doubly degenerate.
The second (third) excited multiplet consists of the state with 
doble- (triple-)particle excitations 
and has $28$- ($56$-)fold degeneracy.
For odd $N$, the multiplets specified by $\{ C \}$ have now different energy. 
All of these results perfectly agree with the fixed-point spectrum 
of the spin-1 two-channel Kondo model \cite{wilson1975},
and the ground state is identified as the LFL state. 

Figure \ref{fig:flow}(b) plots the energy flows for $(J, \Delta) \EQ (0.1, 10.0)$ 
and the results show the behavior of the NFL fixed point for $N \msm2 > \msm2 38$.  
In this case, each multiplet does not change $ \{ C \} $ with $N$, 
and the ground state is now doublet with $\{ C \} \EQ (0,0)$.  
The first excited multiplet has 4 states with $\{ C \} \EQ (\pm1,0)$ 
or $(0,\pm1)$, while 
the second excited multiplet has 10 states with 
$\{C \} \EQ (0,0)$, $(\pm1,0)$, or $(0,\pm1)$.
This agrees with the spectrum at the NFL fixed point \cite{Pang1991}. 
Thus, when the $\Gamma_1$ excitation energy becomes large $\Delta \msm2 \gg \msm2 J$, 
$\Gs$ is effectively decoupled and the system is reduced to 
the ordinary quadrupole Kondo model with 
two symmetric spin channels \cite{Cox1987}.  

\begin{figure}[tb]
  \begin{center}
  \includegraphics[scale=0.37]{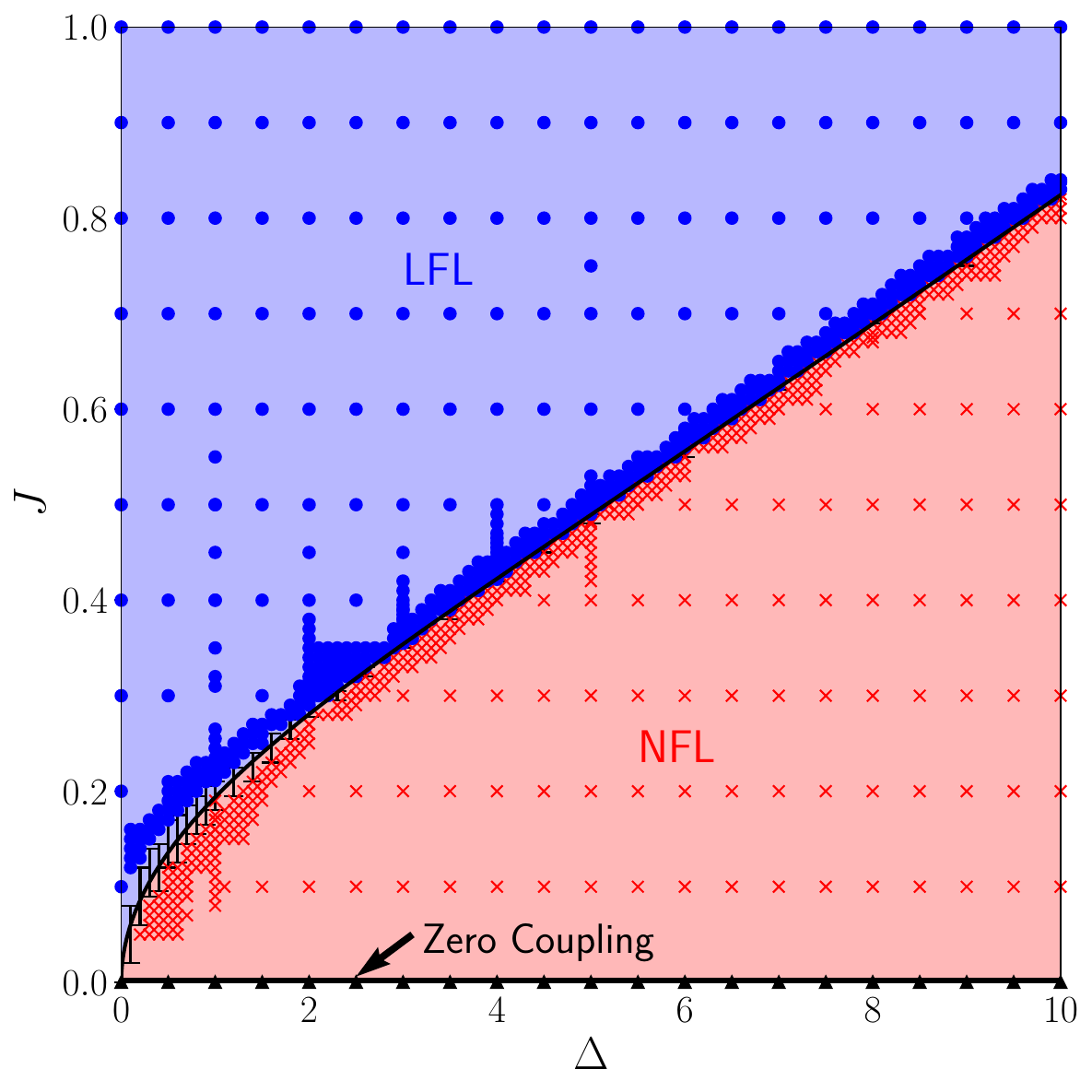}
  \caption{(Color online) 
  The phase diagram of EQKM. 
  The LFL and NFL phases are marked by circles and crosses, respectively.
  }
  \label{fig:phase_diag}
  \end{center}
  \end{figure}

The $J$-$\Delta$ phase diagram is determined following the above procedure 
and the result is shown in Fig.~\ref{fig:phase_diag}.  
The LFL and NFL phases are separated by a smooth boundary $J_c (\Delta)$
that becomes straight at large $\Delta$.  
This can be understood by considering the strong coupling limit, 
which is described by the local term $H_0$. 
We can easily diagonalize it and find a singlet ground state 
for $J/\Delta > \frac{2}{27} = R_c$ 
and doublet ground state for $J/\Delta < R_c$. 
The degeneracy is elevated to triplet at the level crossing point.  
It is clear that the asymptotic form in the large-$\Delta$ region 
is $J_c \sim R_c \, \Delta$.   
The small coupling part (small $\Delta$ and $J$) is most 
interesting, but it is not easy to precisely determine $J_c (\Delta)$
due to slow convergence in the NRG iterations
that is enhanced by a small energy scale. 
The boundary shows a singular $\Delta$-dependence near the origin.  
If this dependence is a simple power 
$J_c \sim \Delta^{\alpha}$, 
the exponent may be determined by 
analyzing the three data points for the smallest $\Delta$ values,  
and the result is
$0.15 < \alpha < 1.4$ despite rough estimate 
due to large error bars.  
For fitting the boundary in the whole range of $0 < \Delta < 10$, 
we have tried several functions and 
the result is quite nice with 
$ J_c = ( a_2 \Delta^2 + a_3 \Delta^3 + a_4 \Delta^4 ) {}^{1/4}$,
where 
$a_2 \EQ 0.0013(1)$, 
$a_3 \EQ 0.58(9) \times 10^{-4}$, and 
$a_4 \EQ 0.27(12) \times 10^{-4} $.   
Its small-$\Delta$ asymptotic form is 
a power with the exponent $\alpha =1/2$, which falls 
in the window determined above. 
Finally, the $J=0$ part of the phase diagram is the free electron phase
where the local impurity is completely decoupled 
from conduction electrons. 

The EQKM has two phases in the parameter space 
even though the system does not change its channel symmetry, 
and this differs from the canonical behavior of the two-channel spin Kondo model. 
In order to explore other novel behaviors, we investigate in detail 
the scaling of the Kondo temperature $\TK (J)$ in the LFL phase 
near the phase boundary. 
To define $\TK$, we follow Wilson's original idea of crossover 
and keep track of the NRG flow of some energy level \cite{wilson1975,Pang1991}. 
The ground state belongs to the subspace of $\{C\} =(0,0)$.
We here choose the energy of the first excited state 
in this subspace and record its energy flow 
$E_1 (N)$ with the NRG step $N$. 
Note that the scaling of $\TK (J)$ does not depend on 
the choice of this eigenstate, 
since all the low-energy states show a common convergence behavior 
in the $N \rightarrow \infty$ limit.
The chosen one is the state that shows 
the simplest convergence behavior even for not large $N$ 
and thus the easiest to analyze. 
We then define the crossover step $N^\ast$ 
by the relation 
$E_1 (N^\ast) = (1 \pm \frac{1}{5} ) E_1^{\infty}$,  
where $E_1^{\infty}$ is the value at the strong coupling fixed point. 
In practice, we fitted $E_1 (N)$ by the function $ E_1^\infty + b_1 \Lambda^{-N/2} + b_2 \Lambda^{-N} $
and obtained a fractional value for $N^\ast$ 
and defined the Kondo temperature by $\TK \EQ t \, \Lambda^{-N^\ast/2}$.  
The energy unit is set to $t =1$ as explained before.  
Figure 3 shows the energy flow of $E_1(N)$ for 101-points in the region $0.010< J-J_c(\Delta)<0.111$. 
All the plots fall on the universal curve within small numerical errors,
and this confirms that the low-energy physics is governed by the same fixed point.
\begin{figure}[tb]
    \begin{center}
        \includegraphics[scale=0.43]{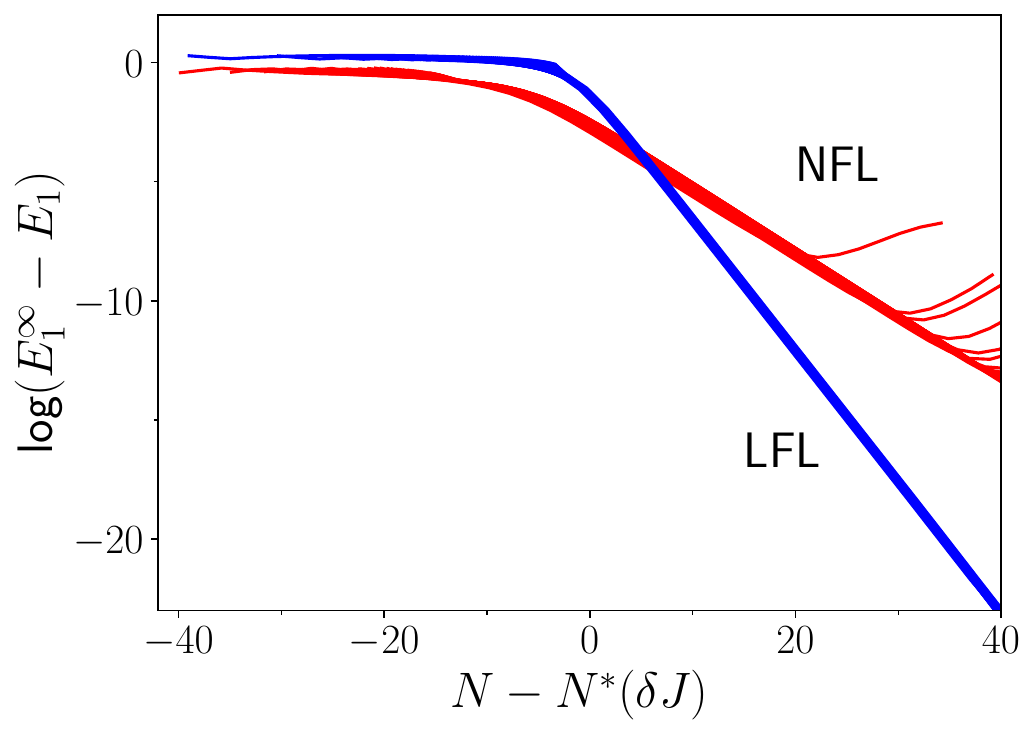}
        \caption{(Color online) 
        NRG energy flows for 101 $J$-points at $\Delta=1.0$ in the LFL region (blue) 
        and for 40 $J$-points at $\Delta=2.0$ in the NFL region (red).
        $\delta J \equiv J - J_c(\Delta)$. 
        The step number is shifted by $N^\ast (\delta J)$.} 
      \label{fig:track}
\end{center}
\end{figure}

\begin{figure}[tb]
\begin{center}
    \includegraphics[scale=0.36]{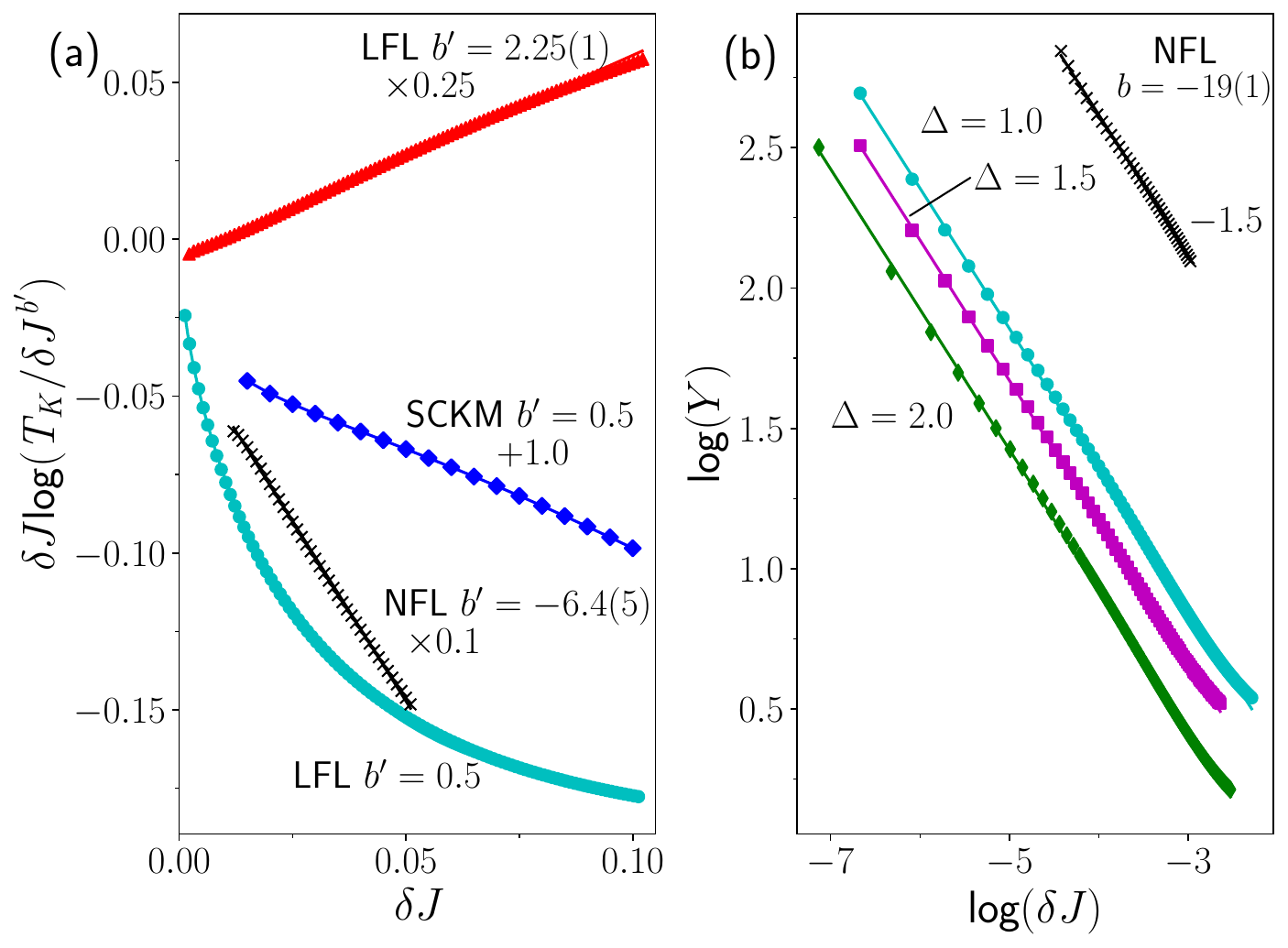}
    \caption{(Color online) Scaling of the Kondo temperature $\TK$.
            (a) $\delta J \log ( \TK / \delta J^{b^\prime} )$ 
            vs $\delta J \EQ J \msm2 - \msm2 J_c (\Delta)$.  
            The results of the EQKM in the LFL and NFL phases are compared with those of the SCKM.  
            Some data are shifted vertically, while some other are multiplied by a factor.
            They are marked by ``$+1.0$" or ``$\times \, 0.25$" etc.
            (b) Double-logarithmic plot of $Y (\delta J)$ for the EQKM.  
            $Y \EQ - \log [ \TK / (A \{J \msm2 - \msm2 J_c (\Delta) \}^b) ]$ 
            using the parameters in Table \ref{tab:fitting}. 
            The slope of the lines coincides with the expected exponent $1/2$ in the stretched 
            exponential.
            The curve of the NFL phase is shifted downwards by 1.5.}
\label{fig:kondo_temp}
\end{center}
\end{figure}

The scaling form of the Kondo temperature is 
$\TK \! \sim \! A' J^{b'} \exp ( - c' /J)$ 
with $b' \EQ 1/2$ for the simplest case,
\textit{i.e.,} the spin-1/2 single-channel Kondo model (SCKM) \cite{wilson1975}. 
Let us first examine whether $\TK$ in the EQKM follows this scaling.  
Note that $J$ should be replaced by the distance 
from the phase boundary $\delta J$. 
Figirue \ref{fig:kondo_temp} (a) shows the scaled form 
$f(\delta J) = \delta J \log [ \TK (J) /\delta J^{b^\prime} ]$ 
for the fixed value $\Delta =1$.  
If the above scaling works, $f (\delta J)$ should show a straight 
line, but the result for $b^\prime=1/2$ is very nonlinear 
and this indicates a different scaling form.  
To check this point, we repeated the same procedure for the SCKM 
to determine its $\TK$ and plotted the result in the same panel. 
This time, the result shows a nice linear behavior confirming 
the expected scaling form for the SCKM. 

We have tried many scaling functions for fitting $\TK (J)$ of the EQKM. 
It turned out that the most promising one was a stretched 
exponential 
$F(\delta J) \exp ( -c/\!\!\sqrt{\delta J} ) $ 
with the prefactor 
$F(\delta J) \EQ A (\delta J)^b $. 
The parameters are optimized 
by minimizing fitting error for $\log \TK$, 
and the results are listed in Table \ref{tab:fitting} for three $\Delta$ values.  
Using these optimized parameters, 
Fig.~\ref{fig:kondo_temp}(b) shows 
$\textstyle Y \EQ - \log [ \TK / F(\delta J) ]$, 
which is expected close to 
$\sqrt{\delta J}/c$.  
If this is the case, the double-logarithmic plot of $Y (\delta J)$ 
should show a straight line with slope $\frac12$, 
and one can see that this works nicely in the panel (b). 
Another promising scaling function  
is an ordinary Kondo form with a generalized  
exponent shown in Fig. \ref{fig:kondo_temp}(a).  
The fitting error is minimized to 0.463 
by using the exponent $b^\prime =2.25(1)$ together with 
$\log A' \EQ 2.59(4)$,  
$c' \EQ 0.0244(2)$, 
and $J_c \EQ 0.210(1)$, 
but this form is unacceptable. 
Recall $1/c'$ corresponds to 
the effective conduction electron density of states, 
and its value seems unphysically large.  
Another reason is that 
the fitting error is more than 3 times 
larger than the case of stretched exponential form.  
These facts strongly support the proposed 
stretched exponential for the best scaling function. 

In order to consider the origin of the stretched exponential scaling, 
it is instructive to recall that this type of scaling also appears at 
the Kosterlitz-Thoules transition \cite{kosterlitz1973}, 
which is governed by two parameters 
(\textit{i.e.}, stiffness $\delta K \EQ K \! - \! K_c$ 
and vortex fugacity $g$).  
Their RG equations have the $\beta$-functions which 
start from the second-order terms, and this is common to 
those of the SCKM with spin anisotropy \cite{Hewson1993}. 
As the bare couplings change with temperature $T$ and 
cross the separatrix line at $T_{\mathrm{KT}}$, 
the scaling function shows 
the factor $\exp (\pm \mathrm{const.}/\sqrt{T \! - \! T_{\mathrm{KT}}})$ 
in the disordered phase.  
Thus, we may expect that the RG equations of multiple 
coupling constants explain the observed scaling behavior 
if their $\beta$-functions have a proper form.  
It is natural to consider three effective coupling constants 
for the present EQKM.  
One is the excitation energy $\Delta$ and the other two are 
quadrupole exchange constants $J_{33}$ and $J_{31}$.  
The impurity quadrupole operators in $H_0$ consist of two parts: 
$Q_{\mu} \EQ Q_{\mu}^{(33)} \msm2 + \msm2 a Q_{\mu}^{(31)}$ 
($\mu \EQ z, x$) 
where $a \EQ \sqrt{35}/2$ is the amplitude 
of $\Gamma_3$-$\Gamma_1$ transition.  
At each RG operation, their exchange couplings 
$J_{33}$ and $J_{31}$ are renormalized 
and the renormalization is generically different between them.   
Construction and analysis of their RG equations is an important 
step for a better understanding of the unconventional scaling in the EQKM. 
However, it is beyond the scope of the present study 
and we leave it for future study.  

Following the same procedure, 
we also analyzed the Kondo temperature 
in the NFL phase at $\Delta=2.0$. 
We once again used the energy of the first excited state $E_1(N)$ 
in the subspace $\{ C \} = (0, 0)$
to determine $\TK$ 
using the same definition \cite{Cox1998}. 
The energy flow is shown in Fig. 3 for 40 points in the region $0.012<\delta J<0.051$.
All the plots once again fall on the universal curve.
This time, however, some plots move away for large $N-N^\ast(\delta J)$, 
but this is due to numerical errors.
In the NFL phase, the ground state is a doublet 
and they appear in the same $\{ C \}$ subspace.
Numerical errors lift this degeneracy and destabilize the RG flows.
Figure \ref{fig:track} shows the asymptotic slope $-0.2473(2)$
different from the value $-1/2$ in the LFL region. 
This indicates the RG eigenvalue $y \sim -1/4$ for the leading irrelevant operator 
around the NFL fixed point.
Figures \ref{fig:kondo_temp}(a) and (b) show that the proposed scalings work quite reasonably.
However, the determined value of the exponents $b^\prime=-6.4(5)$ and $b=-19(1)$, 
seem unphysical.
This may indicate the possibility of another new scaling behavior distinct, 
but we also leave this point for future studies.
\begin{table}[t]
    \centering
    \caption{Optimized parameters including $J_c(\Delta)$ for fitting $\TK$ 
    in the EQKM with a stretched exponential form.}
    \label{tab:fitting}
    \begin{tabular}{ccccc}
        \hline \hline
        $\Delta$ & $\log A$ & $b$ & $c$ & $J_c$ \\ \hline
        1.0      & 2.23(3)  & 1.49(1)   & 0.525(3) & 0.211(1) \\
        1.5      & 2.07(3)  & 1.51(1)   & 0.436(3) & 0.249(1) \\
        2.0      & 1.96(3)  & 1.51(2)   & 0.341(2)  & 0.284(1) \\
        \hline \hline
    \end{tabular}
\end{table}

\begin{figure}[tb]
    \begin{center}
        \includegraphics[scale=0.43]{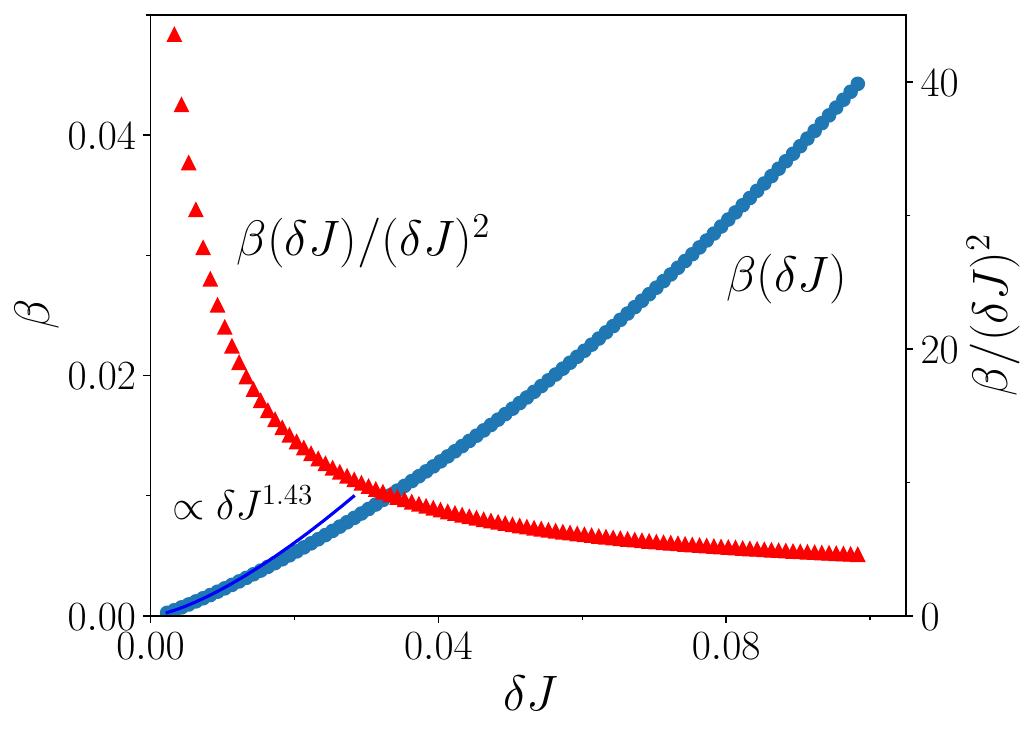}
        \caption{(Color online) 
        $\beta$-function of the EQKM for $\Delta=1.0$. 
        The used $J_c(\Delta)$ value is taken from Table \ref{tab:fitting}.}
        \label{fig:beta}
    \end{center}
\end{figure}
We can further confirm 
the different $\TK$ scaling in the LFL phase of the EQKM 
by analyzing an effective $\beta$-function.  
Suppose $\TK$ is determined by the RG equation of one parameter 
$\delta J (N)$ alone, 
$d \delta J /d N = \beta (\delta J)$, 
as in the case of the SCKM.
Then, the definition $T_K = \Lambda^{-N^{\ast}/2}$ 
leads to the relation 
$\beta \EQ 
(\log \Lambda /2) [d (\log T_K)/\mathrm{d} J]^{-1}$ 
evaluated at $J \EQ J_c \msm2 + \msm2 \delta J$.  
We numerically calculated this and 
the determined $\beta (\delta J)$ is shown in Fig.~\ref{fig:beta}.  
As for the SCKM, it is known that the $\beta$-function 
starts from a marginally relevant term 
$\beta (J) \EQ \rho J^2 \msm2 + \msm2 c_3 J^3 + \cdots$ 
\cite{Anderson1970}.
The determined $\beta (\delta J )$ apparently differs from 
this conventional form, 
as shown by a divergent behavior of $\beta /(\delta J)^2$.  
The small-$\delta J$ part is approximated by a simple power 
$(\delta J)^{\nu}$ with the fractional exponent 
$\nu \EQ 1.43(3)$.  
This nonanalyticity of the effective $\beta$-function 
exhibits a failure of a single RG equation.  
Thus, one needs to consider coupled RG equations of multiple 
coupling constants, which was also anticipated 
from the determined scaling function 
of stretched exponential form.

In summary, we have performed the NRG study on
the quadrupole Kondo effect with taking into account 
the crystal-field $\Gamma_1$ singlet state with excitation energy $\Delta$.  
The determined phase diagram has 
two phases, 
local Fermi liquid for $J\gg \Delta$ 
and non-Fermi liquid for $J \ll \Delta$.  
We have also found a new scaling form of 
the Kondo temperature in the local Fermi liquid phase, 
and it is a stretched exponential function.  
To achieve a full understanding of this behavior,  
we need to analyze 
the RG equations of multiple coupling constants 
and explore a related boundary conformal theory 
as well as perform numerical calculations with higher precision. 
They are important future studies.

{\it Acknowledgements} 
The main part of numerical calculations was performed using 
the facility at the Supercomputer Center, ISSP, the University of Tokyo.

\end{document}